\documentclass[ aip, amsmath,amssymb, reprint ,jcp]{revtex4-1}
\usepackage{graphicx}
\usepackage{dcolumn}
\usepackage{bm}
\usepackage[utf8]{inputenc}
\usepackage[T1]{fontenc}
\usepackage{mathptmx}
\usepackage{etoolbox}
\makeatletter
\def\@email#1#2{%
 \endgroup
 \patchcmd{\titleblock@produce}
  {\frontmatter@RRAPformat}
  {\frontmatter@RRAPformat{\produce@RRAP{*#1\href{mailto:#2}{#2}}}\frontmatter@RRAPformat}
  {}{}
}%
\makeatother
\begin{document}

\title{Hydrogen Bond Topology Reveals Layering of LDL-like and HDL-like Water at its Liquid/Vapor Interface}
\author{Pal Jedlovszky}
\affiliation{Department of Chemistry, Eszterhazy Karoly Catholic University, Leanyka utca 12, H-3300 Eger, Hungary
}
\author{Christoph Dellago}
\affiliation{Faculty of Physics, University of Vienna, Boltzmanngasse 5, Wien A-1090, Austria}
\affiliation{Research Platform on Accelerating Photoreaction Discovery (ViRAPID), University of Vienna, 1090 Vienna, Austria}
\author{Marcello Sega}
\affiliation{Department of Chemical Engineering, University College London, London WC1E 7JE, United Kingdom}
\email{m.sega@ucl.ac.uk}

\date{\today}

\begin{abstract}\noindent{}

The discovery of high-density liquid (HDL) and low-density liquid (LDL) water has been a major success of molecular simulations, yet extending this analysis to interfacial water is challenging due to conventional order parameters assuming local homogeneity. This limitation previously prevented resolving the composition of the surface layer of the liquid/vapour interface. Here, we apply a recently introduced topological order parameter [R. Foffi and F. Sciortino, J. Phys. Chem.
B 127, 378–386 (2022)] to analyze the composition of the water/vapor interface across a broad temperature range. Our results reveal that LDL-like water dominates the outermost region at all temperatures, while HDL-like water accumulates beneath it, presenting a clear layering roughly below the temperature of maximum density. This structured stratification, previously inaccessible, highlights the power of the topological order parameter in resolving interfacial molecular heterogeneity and provides new insights into the structural properties of water at interfaces.
\end{abstract}

\maketitle

\section{Introduction}

Water is characterized by a surprisingly rich phase behavior, and several of its anomalies are often attributed to the existence of two distinct liquid forms\cite{Gallo2016}: high-density liquid (HDL) and low-density liquid (LDL). The concept of liquid polymorphism in supercooled water has been extensively studied, leading to the hypothesis of the existence of a liquid-liquid critical point where HDL and LDL coexist in equilibrium\cite{Poole1992}. This framework is supported by many experimental\cite{sellberg2014ultrafast,nilsson2015structural,dehaoui2015viscosity,kim2017maxima,kim2020experimental,amann2023liquid} and computational\cite{Poole1992,Mishima1998,jedlovszky2005liquid,Palmer2014,Palmer2018,singh2019thermodynamic,sciortino2025constraints} studies, to cite only a few.

At temperatures above the liquid-liquid critical point, fluctuations between HDL-like and LDL-like regions persist, with water existing as a mixture of these two structural motifs.  A key open question in this field concerns the composition of water at its liquid/vapor interface, which is known to have markedly different structural and dynamical properties from the bulk\cite{Partay2008,Sega2014a,bonn2015molecular,fabian2017single}.
Malek and colleagues investigated the distribution of HDL-like and LDL-like molecules in water droplets\cite{Malek2018,malek_surface_2019}, finding that the interface exhibits structural signatures distinct from the bulk. Several simulations of liquid/vapor interfaces in slab configuration confirmed the presence of a compact layer at the interface, in the shape of a shoulder, or apophysis, in the global density profiles \cite{Matsubara2007, Abe2014, Haji-Akbari2014, Haji-Akbari2017, Malek2018, Wang2019,Vins2020}. Recently, we have carried out an investigation of flat liquid/vapor interfaces from the normal to the supercooled regime\cite{gorfer2023high}. While the main aim of our previous work was to investigate the second inflection point in the surface tension of water, we also noticed the accumulation of HDL-like water in the interfacial region for temperatures lower than, approximately, the temperature of maximum density.

However, our previous analysis excluded the outermost molecular layer due to limitations of the order parameter (the fifth-neighbor distance\cite{cuthbertson_mixturelike_2011} criterion) used to distinguish HDL-like from LDL-like water, which assumes a locally homogeneous environment. A variety of order parameters have been proposed to distinguish LDL-like and HDL-like water, many of which have been shown to be largely equivalent in their ability to distinguish the two motifs\cite{foffi2022correlated}. The vast majority of the available order parameters assume an implicit homogeneity in the local molecular environment. However, this assumption is problematic in interfacial settings, which are dominated by strong gradients.

Foffi and Sciortino recently introduced a topological order parameter\cite{foffi2022correlated} that classifies water molecules based on the distance to their fourth hydrogen-bonded topological neighbor. Unlike local structural descriptors, it does not assume a homogeneous environment and remains well-defined at interfaces. A molecule in the outermost layer can still be classified as long as it has a fourth topological neighbor, making this order parameter particularly suited for interfacial studies.

In this work, we apply the topological order parameter to analyze the molecular-layer composition of the water/vapor interface over a broad temperature range. 
This approach enables us to overcome the previous limitations and resolve the composition of the surface layer, showing a previously inaccessible layering phenomenon, where LDL-like water dominates the outermost layer while HDL-like water accumulates just beneath it, forming two alternating layers. 

\section{Methods}
\begin{figure}[!t]
   \centering
   {\includegraphics[trim=20 18 10 10, clip, width=1\columnwidth]{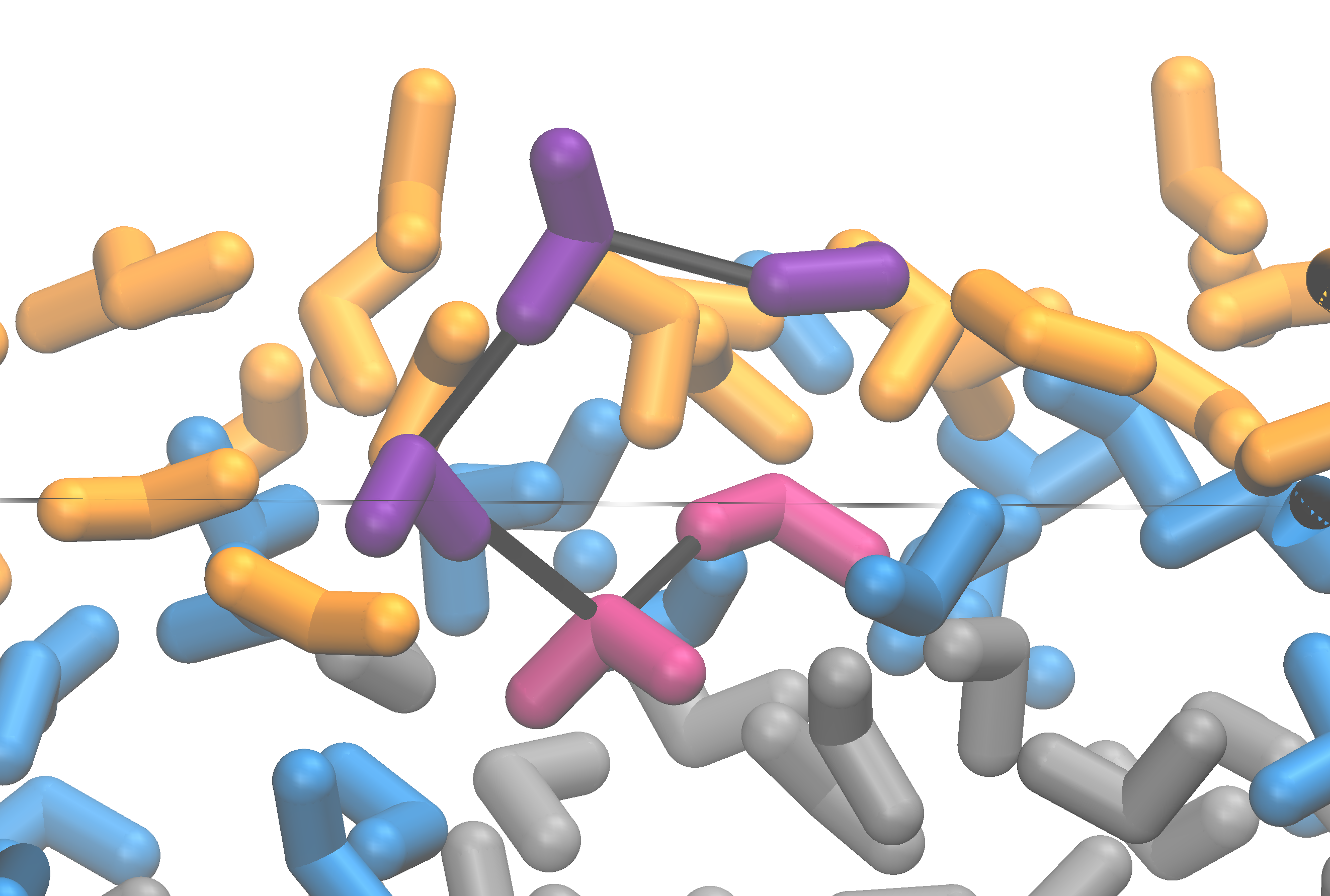}}
   \caption{
   Simulation snapshot (detail) of the interfacial region. Yellow and purple: first layer; Blue and pink: second layer ; gray: other molecules. Purple and pink molecules belong to a chain of 4 hydrogen bonds (black lines), the first and last molecules of the chain being HDL-like.\label{fig:snap}}
\end{figure}

We performed molecular dynamics (MD) simulations using the GROMACS 2019.3 package\cite{Abraham2015} and the TIP4P/2005 water model\cite{Abascal2005}, which accurately captures the phase behavior of water, including the presence of the liquid-liquid phase transition~\cite{Russo2014,Abascal2010,debenedetti2020second}.

We started the simulations from the last equilibrium configuration taken from our previous work\cite{gorfer2023high}. The original setup and equilibration protocol consisted in starting from a 50 × 50 × 50 \AA$^3$ cubic box with 4017 water molecules, initially equilibrated in the bulk phase, from which we generated the slab configuration liquid/vapor interface by expanding the box edge along z-direction to 200 \AA. The rigidity of the water molecules was ensured by using the SETTLE~\cite{miyamoto1992settle} algorithm. Periodic boundary conditions were applied along all directions. All simulations were conducted in the canonical ensemble at temperatures ranging from 198.15 to 348.15 K. The temperature control was maintained using the Nosé–Hoover thermostat\cite{Nose1984,Hoover1985} with a time constant of 2 ps. Note that this is well within the safe operating range for this thermostat\cite{ke2022effects}: a test on the lowest temperature system with a time constant of 0.5 ps yielded a statistically indistinguishable mean and variance of the energy. The equations of motion were integrated using the leapfrog algorithm\cite{allen2017computer} with a timestep of 1 fs. The long-range contributions to energy and forces were computed using the smooth variant of the Particle Mesh Ewald (PME) method for both the electrostatic and dispersion interactions\cite{essmann1995smooth,sega_long-range_2017-2}. For the PME algorithm we used a real-space cutoff of 13 \AA, a Fourier grid spacing of 1.5 \AA{} and a real-space relative contrubution at the cutoff  of $10^{-5}$ and $10^{-3}$ for electrostatics and dispersion, respectively. In our previous work, the systems were simulated at equilibrium for 90 ns, with the exception of the one at 198.15 K, which was simulated for 250 ns. For this investigation, we started from the final equilibrium configurations and added production runs of 80 ns, storing configurations to disk every 20 ps for subsequent analysis. The molecular structure of the liquid/vapor interface was analyzed using the Pytim package\cite{Sega2018}, based on MDAnalysis\cite{Michaud-Agrawal2011, Gowers2016} and available online at \url{https://github.com/Marcello-Sega/Pytim}. We defined the interface using the Identification of Truly Interfacial Molecules (ITIM) algorithm\cite{Partay2008} as implemented in Pytim. ITIM assigns molecules to the interfacial layer if any of their atoms are reachable from the vapor phase using a probe sphere of a certain radius, here chosen to be 1.5 \AA. This approach accurately captures capillary wave fluctuations\cite{buff1965interfacial,stillinger_capillary_1982,chacon2003intrinsic}, allowing for a precise definition of the instantaneous interface. Once identified the interfacial layer, we repeat the procedure for the remaining molecules iteratively, therby assigning the successive molecular layers located at increasing depth into the liquid phase.

The fraction of HDL-like and LDL-like molecules in each layer was later determined using the topological order parameter $\psi$. The order parameter $\psi$ is defined as the distance between the closest of two water molecules that are connected by four hydrogen bonds (i.e., the extremal molecules in a chain of five hydrogen bonded ones, as depicted in Figure\ref{fig:snap}). 

To compute the topological order parameter, we use a graph-based approach. First, we construct the adjacency matrix $M_{ij}$ of molecules $i$ and $j$ that are hydrogen-bonded, where $M_{ij}=1$ if the $i$-th and $j$-th molecules are hydrogen-bonded, and zero otherwise.
A hydrogen bond is identified using a geometric criterion requiring a (donor-acceptor) oxygen-oxygen distance O$_\mathrm{don}\cdots$O$_\mathrm{acc}$ < 3.5 \AA{} and an (acceptor) oxygen-hydrogen distance O$_\mathrm{acc}\cdots$H < 2.5 \AA, roughly corresponding to the (in fact, temperature dependent) first minima of the corresponding pair distribution functions. 

Several alternative approaches including angular or energetic constraints are possible, but they are all essentially qualitatively equivalent (the reader is referred to the excellent comparison of many criteria by Skinner and coworkers\cite{kumar2007hydrogen}).

Starting from the adjacency matrix, we obtain the shortest paths (in terms of number of bonds) between all molecular pairs using Dijkstra’s algorithm\cite{dijkstra1959note} and limiting the exploration of nodes in the connectivity graph to topological distances of at most four links. The molecular pairs with a topological distance of exactly four are extracted, and their Euclidean distances are computed using the minimum image convention. For each molecule, we determine its closest fourth topological neighbor by selecting the pair with the smallest Euclidean distance, which represents the order parameter $\psi$.

\section{Results and Discussion}\label{Ch:Results}

To investigate the structure of interfacial water, we define the density profiles across the slab of HDL-like and LDL-like molecules based on the topological order parameter $\psi$, following the approach introduced by Foffi and Sciortino. A molecule is classified as HDL-like if $\psi<5$\AA{}, and as LDL-like otherwise. The choice of this threshold is motivated by the apparent separatrix in the bimodal distributions of $\psi$ reported in Refs.\cite{foffi2022correlated} and \cite{foffi2024identification}. The two peaks corresponding to HDL-like and LDL-like environments are centered around 3.5 and 6.5\AA{}, respectively. The threshold at 5\AA{} represents a natural separation between these two states, effectively distinguishing the two molecular environments.

\begin{figure}[!t]
   \centering
   {\includegraphics[trim=10 0 40 35, clip, width=1\columnwidth]{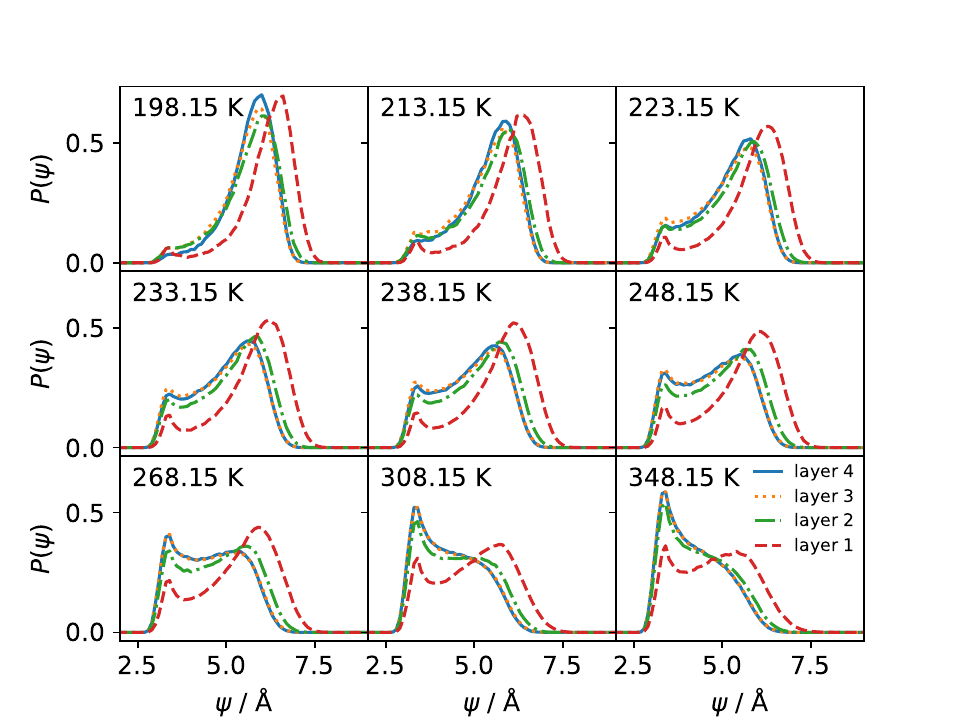}}
   \caption{
   Probability density distributions of the topological order parameter $\psi$ in the first 4 layers at selected temperatures. \label{fig:distribution}}
\end{figure}

Figure~\ref{fig:distribution} shows the probability density distribution of $\psi$ for a selected set of temperatures and for the first 4 surface layers. Note that while the location of the characteristic peak of HDL-like water at around $\psi = 3$ \AA{} shows no dependence on temperature or layer, the LDL-like water peak location at larger values of $\psi$ (between 5.5 and 7 \AA{}) does indeed have a mild temperature dependence and noticeable shift of approximately 0.5\AA{} when moving from the first to the subsequent layers. The distributions in layers 2-4 differ only marginally and show that water becomes essentially homogeneous at the structural level already from the second or third layer. The fact that the peak at larger values of $\psi$ has a systematic shift underlines a strutural difference in the surface layer only for molecules that are LDL-like.

\begin{figure}[!t]
   \centering
   {\includegraphics[trim=16 18 10 10, clip, width=1\columnwidth]{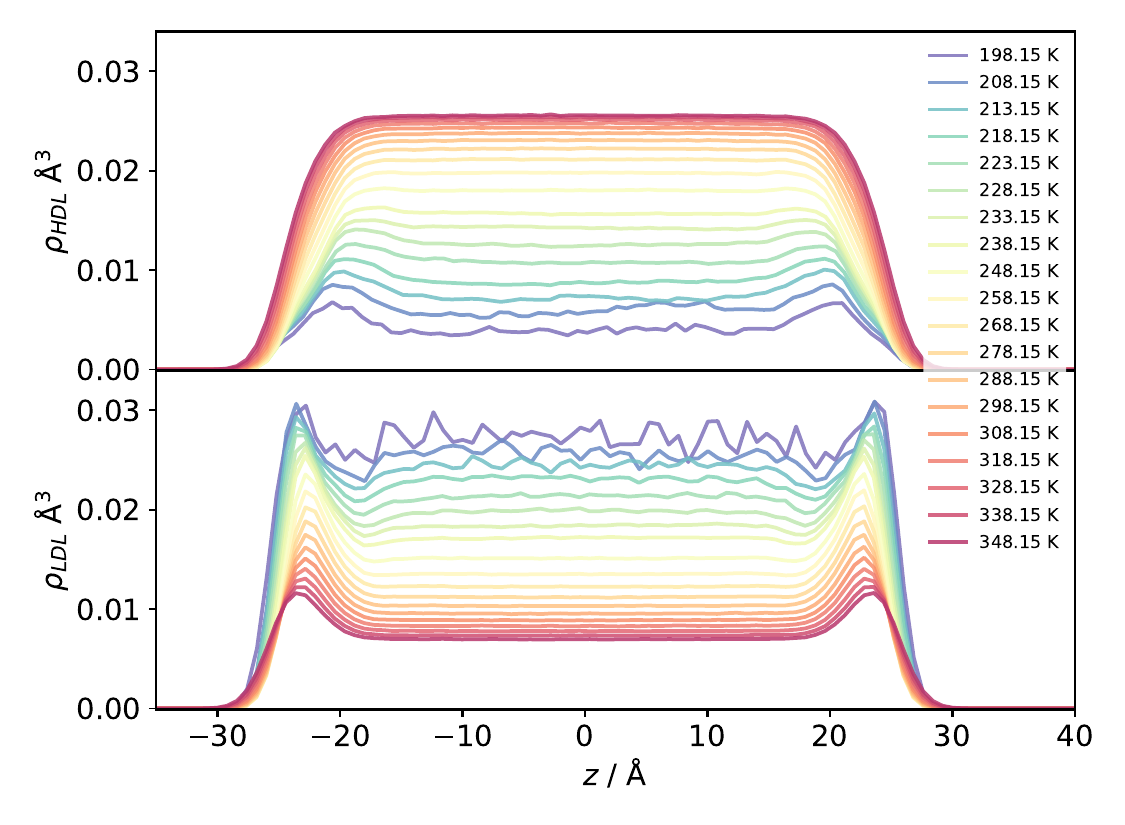}}
   \caption{
   Number density profiles of HDL-like and LDL-like water molecules across the slab. \label{fig:rhoHDL_LDL}}
\end{figure}

Figure \ref{fig:rhoHDL_LDL} reports the number density profiles of HDL-like (top panel) and LDL-like molecules (bottom panel) across the liquid-vapor interface. In the bulk region, the relative densities follow the expected temperature trend that sees the concentration of LDL-like water in the bulk increasing upon cooling, especially below the temperature of maximum density. Closer to the interface, we observe a more complex behavior. The HDL-like concentration increases in a subsurface region before decaying toward the interface. This behavior is identical to what we observed in our previous work\cite{gorfer2023high} using the fifth neighbor distance as an order parameter. This is not surprising, as the order parameters used to distinguish LDL from HDL-like water are quite robust\cite{foffi2022correlated}. Using the fifth neighbor distance criterion, however, the concentration and fraction of HDL-like water molecules next to the surface is necessarily a lower bound (and, conversely, the LDL-like is an upper bound).
With the topological order parameter, instead, we are, for the first time, able to probe in a meaningful way the composition at the boundary of the liquid phase.

Going from the bulk towards the surface, the accumulation of HDL-like molecules observed at low temperatures is followed by a layer where the concentration of LDL-like molecules increases. 
However, while the peak in the number density profile of HDL-like molecules disappears at high temperatures, this is not true for the concentration of LDL-like molecules, which are always preferentially found at the interface with respect to the bulk.

The locus of the maximum in the LDL density profile in the bottom panel of Figure~\ref{fig:rhoHDL_LDL} exhibits a narrowing that has a minimum around 280K -- confirming what we obseverved in our previous work\cite{gorfer2023high} -- which concides with the density maximum of the TIP4P/2005 model and is far away from the temperature range of several anomalies associated with the Widom lines\cite{Gallo2016,martelli2019unravelling}.

\begin{figure}[!t]

   \centering

   {\includegraphics[width=1\columnwidth]{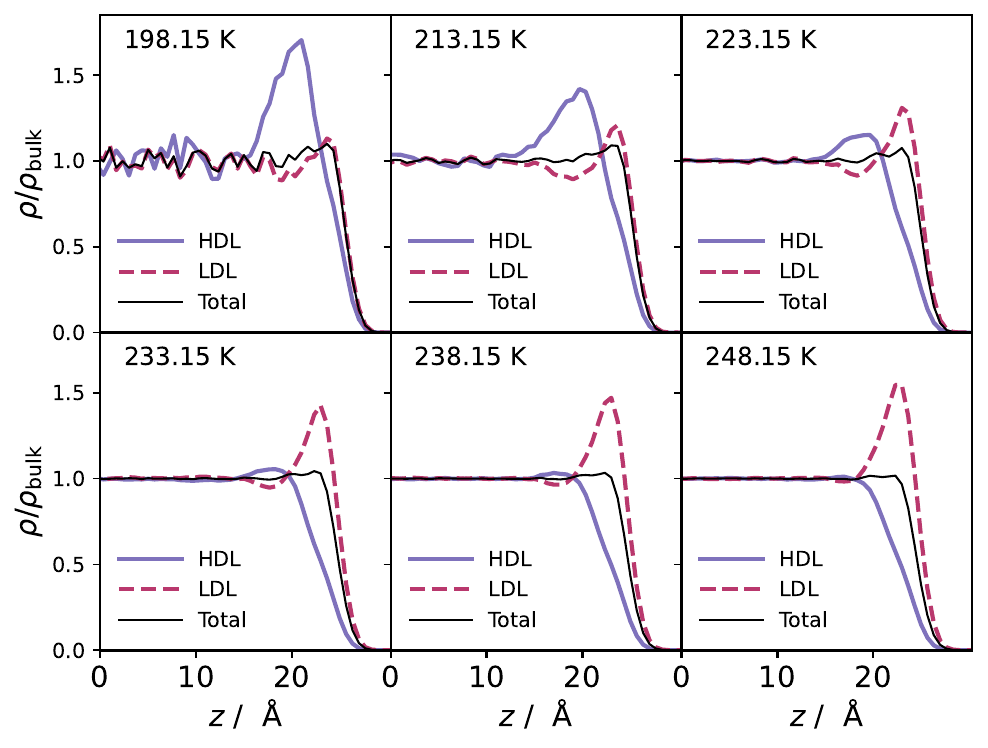}}

   \caption{Number density profiles of HDL-like and LDL-like water molecules (across half of the slab, for clarity) normalised with respect to their bulk values (selected temperatures). \label{fig:layering2}}

\end{figure}

From the position of the peaks one can see that LDL/HDL-like water layering is taking place at low temperatures. 
This can be better seen if one overlays the two profiles, normalising the values to the respective bulk one, as shown in Figure \ref{fig:layering2}.

This normalization highlights the relative variations in concentration and illustrates the emergence of two layers of HDL-like and LDL-like water near the interface.

In the outermost region, where the profiles are dropping to zero, that of LDL-like molecules always follows or is larger than the total one, meaning that LDL-molecules are always present at the surface in larger concentrations than in the bulk for all systems. A local minimum emerges in the LDL-like profile, just below the apophysis. The normalized HDL-like number density profile, instead, is always smaller than the total one in the outer interfacail region, showing a systematic depletion of HDL-like molecules. Below the temperature of maximum density, however, a concentration maximum starts developing at the location of the local minimum of the LDL-like one, suggesting that this is happening in a sub-surface layer. This maximum keeps growing upon further lowering the temperature, eventually developing a large (relative) accumulation of HDL-like water near the surface. The normalized number density profiles clearly show the appearance of alternating layers of increased/decreased concentration (with respect to the bulk value) of HDL-like and LDL-like water molecules. This results highlights the unexpected presence of areas for preferential accumulation of the two structural motifs near the interface.

At low temperatures, the most striking feature is an enhanced layering, where LDL-like water dominates the outermost interfacial layer, while HDL-like water accumulates just beneath it. In correspondence of the LDL-like accumulation one can notice a small shoulder, or apophysis in the total profile that lasts up to temperatures close to the density maximum one\cite{Wang2019,gorfer2023high}. At higher temperatures, both the apophysis and the HDL-like concentration maximum disappear, and only LDL-like surface enrichment is observed, in concomitance with surface depletion of HDL-like molecules.

This behavior is consistent with the previously observed structuring of water\cite{Wang2019,malek_surface_2019,gorfer2023high,Malek2018} at interfaces but is now resolved in terms of its local hydrogen-bond network topology. At low temperatures,  the presence of two alternating layers, one with HDL-like water excess and and one with LDL-like water excess,  suggests a collective reorganization of the hydrogen-bond network near the interface, possibly linked to the mechanisms underlying water’s liquid-liquid transition and the availability of free volume at the liquid/vapor interface.

The persistence of LDL-like water at the outermost molecular layer across all temperatures is particularly noteworthy. This result implies that water at the interface tends to favor the LDL-like local structure even at high temperatures, whereas the enhancement of HDL-like water at the subsurface region is most pronounced at lower temperatures.

The topological order parameter has primarily been applied in bulk water studies, raising the question of whether its application to interfacial systems might introduce systematic biases. To check this, we analyzed a control system in which we artificially removed molecules that are farther than 15 \AA{} from the middle of the slab, leaving a 30 \AA{} thick artificially flat interface. We processed this artificial system identically to the initial one. This approach isolates any potential artifacts arising from the order parameter itself, allowing us to determine whether the observed layering is an genuine feature of interfacial water or a more general consequence of structural correlations in bulk liquid water.

\begin{figure}[!t]

   \centering

   {\includegraphics[width=1\columnwidth]{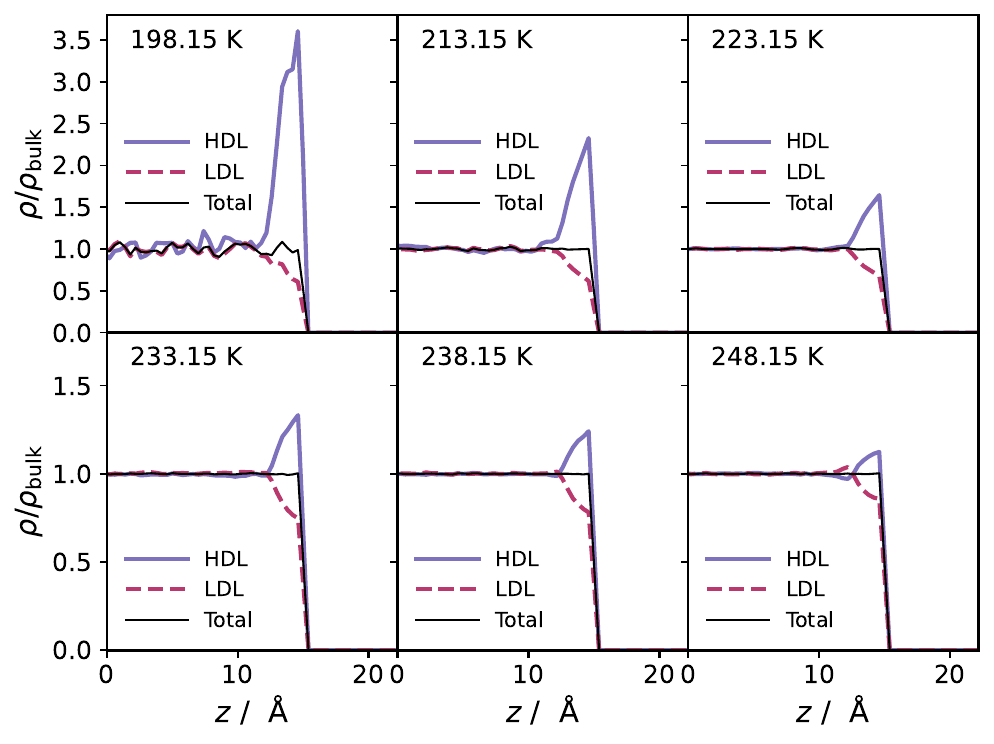}}

   \caption{Number density profiles of HDL-like and LDL-like water molecules in an artificially truncated bulk region (selected temperatures). \label{fig:layeringCut2}}

\end{figure}

Figure~\ref{fig:layeringCut2} presents the HDL-like and LDL-like number density profiles obtained in the artificially truncated bulk region. Maybe unexpectedly, the profiles are not flat. Unlike at the real interface, where LDL-like molecules dominate the outermost layer, and HDL-like molecules accumulate in a subsurface region, at the artificially imposed boundary, there is always a systematic accumulation of HDL-like molecules and a depletion of LDL-like ones.

The behavior observed at the artificial interface can be understood by considering the nature of the topological order parameter $\psi$. Consider two LDL-like molecules at the artificial boundary close to each other and initially at the (shortest) topological distance of three or two links. If one of these links is severed by the cut, but the pair is still connected via another path, this must necessarily have an equal or larger topological distance. Therefore, the two molecules have a chance to be promoted to fourth-bonded neighbors and tagged as HDL-like. The opposite can also happen when a path of four links between two HDL-like molecules is severed. If this is the only path of four links connecting them, they will be tagged as LDL-like. The balance between these two possibilities will depend on the statistics on the number of paths and the state point. The result could be either more or less HDL-like molecules at the artificial cut, depending on the relative number of paths of different lengths, and the balance favors HDL-like molecules in practice.

This explanation provides a strong argument reinforcing the validity of our results at the real water/vapor interface. If the layering observed in Figures~\ref{fig:rhoHDL_LDL} and~\ref{fig:layering2} were merely an artifact of the order parameter, we would expect HDL-like molecules to accumulate at the real surface as well, following the same bias seen in the truncated bulk slab. Instead, we observe the opposite trend: LDL-like molecules accumulate at the real interface, while HDL-like molecules preferentially occupy subsurface layers. This result suggests that the alternating HDL-like and LDL-like layering at the interface is a genuine physical feature rather than a spurious effect of the order parameter.

Furthermore, the presence of alternating layers in the real interfacial system but not in the control system indicates that the structuring of water near the interface is not merely a consequence of cutting the system but rather arises from molecular-scale interactions specific to the liquid-vapor boundary. These results confirm that the topological order parameter is a robust tool for analyzing interfacial water composition and reinforces our previous conclusions.

One further comment is in order. The pronounced peak in the low temperature HDL-like profile at the artificial interface, visible in Figure~\ref{fig:layeringCut2}, is partially a consequence of the normalization to bulk values. As shown in the non-normalized plots in the Supplementary Material, the excess of HDL-like molecules is compensated by a corresponding depletion of LDL-like ones. Since the artificial conversion of LDL-like to HDL-like molecules (via the path truncation mechanism discussed above) is proportional to the local density of LDL-like molecules, the resulting boundary effect is more pronounced at low temperatures, where LDL-like water dominates in the bulk.

It is also important to note that the HDL-like peak in the artificial system appears exactly at the interface, coinciding with the LDL-like depletion. In the natural interface, by contrast, the HDL-like peak consistently emerges one layer behind the surface, following an LDL-like maximum. This shift is also visible in the layer-by-layer profiles and confirms that the accumulation of HDL-like water at the artificial boundary is a direct consequence of the truncation.

Our analysis of the accumulation and depletion layers has been, so far, necessarily qualitative, because the presence of thermal capillary wave fluctuations complicates the interpretation of the number density profiles. In profiles such as those shown in Figures~\ref{fig:rhoHDL_LDL} and~\ref{fig:layering2}, molecules at the same elevation relative to the mean interface position may belong to different instantaneous molecular layers due to surface capillary waves. This issue leads to a smearing effect, where properties of neighboring layers mix, making it challenging to resolve finer structural details at the molecular scale.

To overcome this issue, we analyzed the fraction $x_\mathrm{HDL}$ of HDL-like molecules in each layer as identified by the ITIM algorithm. This approach allowed us to investigate how HDL-like water is distributed on a per-layer basis without interference from capillary fluctuations.

\begin{figure}[!t] \centering {\includegraphics[trim=15 15 10 10, clip, width=1\columnwidth]{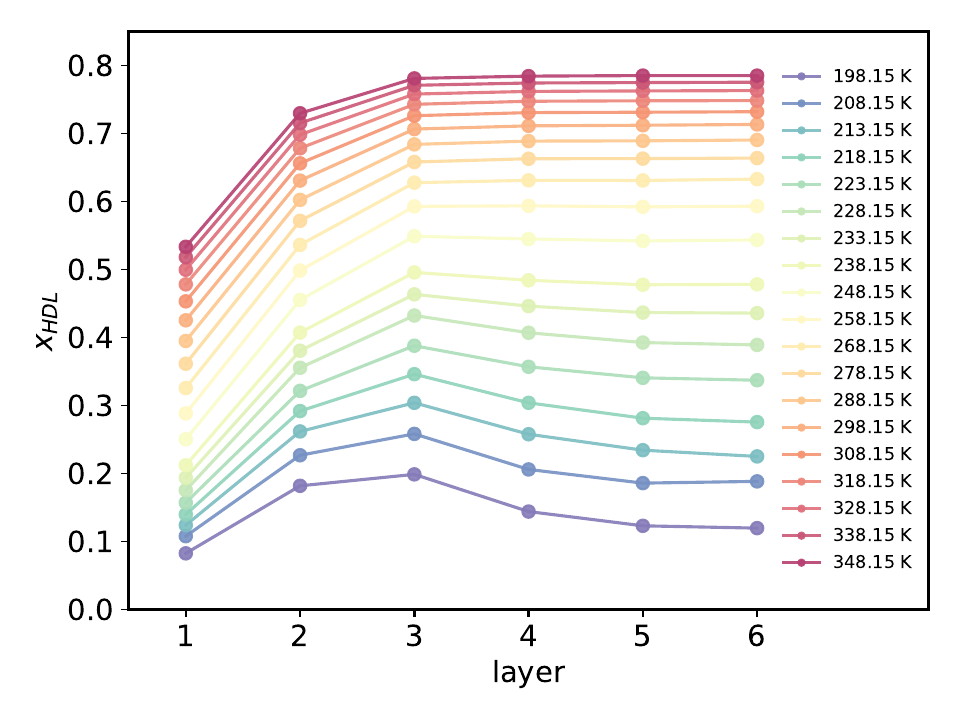}} \caption{Fraction of HDL-like molecules as a function of layer index for different temperatures. The layers are labeled starting from the interface (layer 1).\label{fig:xHDL_layers}} \end{figure}

Figure~\ref{fig:xHDL_layers} shows the fraction of HDL-like molecules, $x_\mathrm{HDL}$ resolved on a molecular layer basis. At low temperatures, we observe that HDL-like molecules accumulate predominantly in the second and third layers rather than in the outermost one. At the lowest temperature of 198.15 K, the fraction of HDL-like molecules in the third layer is about 30\% higher than in the bulk. 

Above 248.15 K, the accumulation of HDL-like water disappears and the fraction of HDL-like molecules across all layers flattens out.

Interestingly, the first layer has consistently the lowest fraction of HDL-like molecules across all temperatures. This confirms that LDL-like water is always dominant in the outermost molecular layer, reinforcing our previous result that the water/vapor interface preferentially stabilizes LDL-like structural motifs. In contrast, at low temperatures HDL-like water accumulates just beneath the interface, creating a structured interfacial layering.

\section{Conclusion}

We have applied the topological order parameter introduced by Foffi and Sciortino\cite{foffi2022correlated} to analyze the molecular-layer composition of the water/vapor interface in terms of the concentration of local LDL-like and HDL-like molecules, revealing a stratification of the local structural motifs. Our analysis shows that the outermost interfacial layer is consistently dominated by LDL-like water. In contrast, HDL-like water is generally depleted from the surface at high temperatures. However, it accumulates mainly in the second and third subsurface layers when the temperature drops below that of maximum density, with the layering becoming more pronounced upon cooling. By repeating our analysis on an artificially generated interface we confirmed that the layering effect is due to a genuine structural property of interfacial water rather than an artifact of the order parameter.

Our results demonstrate the applicability of the topological order parameter to interfacial systems, extending its use beyond bulk water, and provide further insight into the structural heterogeneity of interfacial water. We observed a preference for LDL-like local environments in the outermost molecular layer of the water/vapor interface, something that was not possible with conventional order parameters. We confirm the previously observed accumulation of HDL-like water molecules upon lowering the temperature (a trend opposite to the bulk), locating it in the second and third molecular layer. In general, the presence of alternating layers of HDL-like and LDL-like water suggests a collective reorganization of the hydrogen-bond network near the surface, which may be related to the mechanisms underlying the liquid-liquid transition in supercooled water. 
Further investigations avenues include understanding how this behavior depends on external parameters including mechanical tension or electric field, or how it manifests at different interfaces, including those with hydrophobic liquids, or in confinement. Of particular importance are the dynamical properties of interfacial water. Single particle dynamics via calculation of the diffusion coefficient calculation within a layer, involving tracking when molecules entering and leaving the layer\cite{fabian2017single}, as well as collective layerwise dynamics in terms of hydrodynamic modes\cite{malgaretti2023surface} or effective layer viscosity\cite{jedlovszky2024surface} are possible means to explore the influence of HDL-like and LDL-like molecules on the dynamical properties of water.

\section*{Supplementary Material}
HDL-like, LDL-like and total number density profiles at selected temperatures, not normalized, for the natural and artificially truncated systems.

\acknowledgments

The authors acknowledge the use of the SIGH cluster at the UCL Chemical Engineering Department.
P.J. acknowledges support from NKFIH in the frame of the National Research Excellence Program under project No. 152095. C.D. ackowledges funding by the Austrian Science Fund (FWF) 10.55776/F81.


%

\end{document}